\documentclass[runningheads,anonymous]{llncs}
\usepackage[T1]{fontenc}
\usepackage{defs}

\begin{document}
\title{Non-Termination Proving: 100 Million LoC and Beyond}
%
%\titlerunning{Abbreviated paper title}
% If the paper title is too long for the running head, you can set
% an abbreviated paper title here
%
\def\orcidID#1{\href{http://orcid.org/#1}
    {\protect\raisebox{-1.25pt}{\protect\includegraphics{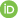}}}
  }
  
\author{Julien Vanegue\inst{1,2}\orcidID{0009-0006-7927-3205} \and
Jules Villard\orcidID{0000-0001-8637-0712} \and 
Peter O'Hearn\inst{3,4}\orcidID{0000-0001-8730-5496} \and
Azalea Raad\inst{2,1}\orcidID{0000-0002-2319-3242}}
\authorrunning{J.Vanegue et al.}
% First names are abbreviated in the running head.
% If there are more than two authors, 'et al.' is used.
%
\institute{Bloomberg, New York, USA \and
Imperial College London, UK \and
Meta FAIR, London, UK \and
University College London, UK}

\maketitle              % typeset the header of the contribution


\begin{abstract}
We report on our tool, \pulseinf, that uses proof techniques to show non-termination (divergence) in large programs. 
\pulseinf works \emph{compositionally} and \emph{under-approximately}: the former supports scale, and the latter ensures soundness for proving divergence. 
Prior work focused on small benchmarks in the tens or hundreds of lines of code (LoC), and scale limits their practicality: a single company may have tens of millions, or even hundreds of millions of LoC or more. We report on applying \pulseinf to \emph{over a hundred million lines} of open-source and proprietary software written in C, \C++, and Hack, identifying over 30 previously unknown issues, establishing a new state of the art for detecting divergence in real-world codebases.
\end{abstract}

\section{Introduction}
\label{sec:intro}

Program non-termination (divergence) is a computationally undecidable problem more difficult than safety: neither it nor its complement, termination, is recursively enumerable~\cite{turing1936computable}. While, assuming unbounded memory,  it cannot be solved on the nose by an algorithm,
there is still the possibility that proving divergence, with acceptable false negative or positive rates, might be done practically via approximate methods. In this paper, we report on a tool, \pulseinf, that has been applied to more than 100M LoC of open source and proprietary software written in C, \C++, and Hack (a typed variant of PHP), identifying more than 30 previously unknown divergence issues,  with 203 false positives.

The basic ideas underlying \pulseinf are intuitively simple. 
First, recall that a sound way to prove divergence in concrete semantics is to look for a state $s$ and a non-empty execution sequence
which circles back to $s$; given such an $s$, which we call a {\em repeating state}\/, we can return to it over and over in an infinite execution. 
%In a Turing machine,  $s$ would contain the contents of the tape, the  head pointing to a cell, and the control state (sometimes just called the state). 
The concrete proof method which searches for repeating states is sound but incomplete and limited. But, when we transport it to abstract semantics, where a single state may denote a set of concrete states, it is very powerful:  this proof method of searching for repeating abstract states is complete, relative to an oracle for the abstract states as sets of concrete states~\cite{UNTER}. Looking for repeating states in an abstract semantics is what \pulseinf does. 

Second, \pulseinf works \emph{compositionally}, following the approach pioneered by \infert~\cite{biabduction-journal}. 
\pulseinf uses separation logic~\cite{reynolds,OHearn19} to describe abstract states, and uses the bi-abductive symbolic execution technique of \infert to allow pre- and post-conditions to be computed from code, with pre/post pairs stored as summaries for procedures. This leads to scalability in the same way as \infert~\cite{FB-CACM}. As \pulseinf is implemented over \infert, it inherits its on-demand,  compositional analysis scheduling algorithm.

%% Finally, in order to determine what errors to report, \pulseinf  restricts attention to functions that are entry-points to a codebase (\eg \texttt{main()}. This stops us from reporting inter-procedural bugs repeatedly; further, these entry points are where we would find a way to trigger a non-termination bug, which could (for example) lead to a denial of service attack. 

The intuitions behind \pulseinf are indeed this simple, but for one hitch: the
``non-empty execution sequence
that circles back to $s$''  must be \emph{under-approximated}, rather than over-approximated to soundly prove non-termination (avoiding false positive non-terminating claims). 
As such, \pulseinf relies on a variant of \emph{incorrectness logic}~\cite{il,isl}, an under-approximate cousin of Hoare logic.
The formal foundations are underpinned by the \unter logic in the precursor paper~\cite{UNTER}, and we do not repeat them here.
Since the publication of \unter~\cite{UNTER}, we have considerably developed the \pulseinf tool further.

Specifically, we have removed sources of false positives and further added a check for \emph{mutually-recursive divergence} (as well as one for infinite loops in the earlier version). 
We have evaluated \pulseinf on large codebases, we have used it industrially on over 50M LoC in Bloomberg and Meta, and evaluated it on over 50M LoC of open source code. 
In total, we have analysed \emph{over a hundred million} lines of code of C, \C++, and Hack programs  and identified more than 30 unique and previously unknown divergence issues. 
We provide a reusable and verifiable evaluation totalling \emph{over fifty million lines of major open source projects}, analysing projects such as the Linux kernel, Wireshark, Bitcoin Core, OpenSSL, SQLite, and many more. 
Our analysis is especially relevant as the number of CVE reports mentioning divergence keeps growing~\cite{UNTER}, and \pulseinf is the first divergence prover capable of scaling to the ever-growing software industry.

In describing our evaluation  we concentrate on open source code, but we also remark on the industrial experience. 
Our title pays homage to the landmark paper on Symbolic Model Checking~\cite{1020}, which established a  state of the art for scalability of model checking that impacted practice.

%
%
%A new theory of under-approximate non-termination proving UNTER~\cite{UNTER} now tackles this undecidable problem by providing a program logic with \emph{No False Positive} by construction. UNTER comes with soundness and complete proofs, as well as an integration with separation logic~\cite{reynolds}, allowing a scalable and compositional implementation of non-termination proving in the \pulse incorrectness separation logic prover~\cite{il}~\cite{isl}~\cite{pulsex} of the \infert toolchain. We provide the first implementation of UNTER theory as an extension to \pulse,  naming this new extension \pulseinf. 
%
%\pulseinf is capable iof identifying non-termination issues due to loop divergence, as well as non-termination due to recursive function divergence. 

\section{Overview of \pulseinf}
\label{sec:overview}

%% \poh{I would love this section to show:
%% \begin{enumerate}
%% \item An inter-procedural bug and summaries and how we report only from main()
%% \item An abstract state which is a looping abstract state. If this were the abstract state for which no single concrete state could do the same job that would be lovely.
%% \item examples as you have below.
%% \item I don't think we need any proof rules.
%% \end{enumerate}
%% }

We summarise the contribution of our tool, \pulseinf, as the first implementation of the \unter~\cite{UNTER} theory for under-approximate divergence proving for loops, as well as a new technique for detecting divergence in (mutually) recursive functions.
%\vspace{2mm}
%\julien{Azalea: Should we show a subset of rules of \unter before transitioning to the below?}
%\vspace{2mm}
%
We divide the divergent code we analyse into two categories,  as represented in \cref{tab:overview}.  
The first category is that of infinite \emph{loops}, whether \emph{unstructured} (\textbf{a}) or \emph{structured} (\textbf{b}). The second category is that of infinite recursions (\textbf{c}). 
For illustrative purposes, we also include examples of terminating programs using unstructured (\textbf{d}) or structured (\textbf{e}) loops, as well as for recursion (\textbf{f}). We expand on these categories and provide a more comprehensive classification of divergence vulnerabilities in Appendix~\ref{sec:cve}.

\begin{table}[t]
    \centering
    \begin{tabular}{|l|l|l|}
        \hline
        \tt previous: x++  &\tt   int x = 1, y = 0; &\tt  f (x) \{ \\
        \tt goto previous; &\tt   while (x) y++;    &\tt  ~~f(x + 1) \}\\
        \multicolumn{1}{|c|}{\textbf{(a)}} &\multicolumn{1}{c|}{\textbf{(b)}} &\multicolumn{1}{c|}{\textbf{(c)}} \\
        \hline
        \tt goto next;  &\tt  int x = 0;          &\tt  f(x) \{ \\
        \tt next: x++   &\tt  while (x < 10) x++; &\tt  ~~if (x < 10) f(x+1) \} \\
        \multicolumn{1}{|c|}{\textbf{(d)}} & \multicolumn{1}{c|}{\textbf{(e)}} & \multicolumn{1}{c|}{\textbf{(f)}} \\
        \hline
    \end{tabular}\vspace{5pt}
    \caption{Examples of non-terminating (\textbf{a}, \textbf{b}, \textbf{c}) and terminating (\textbf{d}, \textbf{e}, \textbf{f}) programs.}\vspace*{-20pt}
    \label{tab:overview}
    \vspace{-5pt}
\end{table}

We have identified previously unreported divergence vulnerabilities in a number of large projects totalling hundreds of millions of lines of code, including both open source and proprietary codebases. Several of these issues have been fixed by developers and others have been reported for triage and prioritisation. In some cases, divergence detection even allowed us to pinpoint weaknesses of the code where classes were identified as \emph{thread-unsafe} as a result of our analysis; see \cref{sec:examples} for more details. 

\paragraph{Unstructured loop divergence} Unstructured control flow is a common source of unintended non-termination in programs. Unstructured control flow is notoriously difficult to analyse, particularly in the presence of \texttt{goto} statements within loops. \pulseinf is capable of detecting \emph{infinite} \texttt{goto} \emph{loops} (\eg (\textbf{a}) in \cref{tab:overview}, where the repeating abstract state is the assertion $\mathit{true}$ at label \texttt{previous}), while determining that other \texttt{goto} statements (\eg (\textbf{b}) in \cref{tab:overview}) do not lead to loops.
While it is most common to have \texttt{goto} divergence on back edges of the program, this is not always the case and we do not identify back-edges as non-terminating or front-edges as terminating. Instead, we rely on the \unter~\cite{UNTER} reasoning rules for detecting non-termination.

\paragraph{Structured loop divergence} Non-terminating loops are by far the most common non-termination bugs found in real-world programs nowadays, as the number of loops in a program is typically much larger than the number of gotos or the number of recursive functions. As such, loop non-termination is a denser property to check for than other non-termination instances. We present simple examples of a finite loop (\textbf{e}) and an infinite loop (\textbf{b}) in \cref{tab:overview}. The repeating abstract state for (\textbf{b}) is the assertion $x == 1$, an assertion that denotes many states including all instantiations of $y$, illustrating the difference from concrete repeating states.
Thanks to an under-approximation of widening at loop iteration, we can symbolically explore loop paths up to a chosen $k$ bound following the intuition that most bugs are shallow. 
In practice, increasing $k$ to a large number only moderately increases the number of alerts, while proportionally increasing the required runtime. 
Our non-termination benchmarks show that running \pulseinf with $k=3$ identifies 14 of the 15 bugs identified with $k=20$. On real programs, \pulseinf on OpenSSL triggers 12 alerts with $k=3$ (1m26 analysis time), while it triggers 21 alerts for $k=10$ (1m37 analysis time), and 27 alerts for $k=20$ (2m37 analysis time). Other targets show a similar trend.

\paragraph{Recursion divergence} Recursive functions are typically used when an inductive data structure is inspected; this is common in XML, JSON and other implementation of standard parsing libraries. 
Uncontrolled recursion can be caused by functions that do not correctly model their exit conditions and call themselves on the same parameters, or by altgother-unintended recursion or mutual recursion caused by programming error.  
We present an example of simple divergent recursion in (\textbf{c}) as well as a terminating one in (\textbf{f}) of \cref{tab:overview}. \pulseinf supports detecting infinite recursion of both recursive and mutually recursive functions, while discarding finite recursions as safe.

\paragraph{Divergence in practice} While for the purpose of this paper we reason in terms of identifying \emph{divergent} programs, actual programming languages runtimes may not always produce non-terminating programs in these cases. Indeed, infinite recursion often causes stack overflow exceptions (\eg in Hack), and infinite loops may cause unpredictable undefined behaviour rather than guaranteed non-termination, for example. The precise form these bugs take is not important for our study.

\section{Examples}
\label{sec:examples}

%We present several examples of divergence bugs found by \pulseinf. 
We have identified over 30 divergence bugs in open-source projects as detailed in \cref{sec:evaluation}.  Here we present several representative examples of divergence bugs found by \pulseinf in real codebases to show the practical applicability of \pulseinf to a wide range of programs.

\begin{figure}[t!]
\hrule
\begin{lstlisting}[numbers=left,basicstyle=\small]
static char *svg_dump_path(SVG_PathData *path){
    u32 i, *contour;  contour = path->contours;
    (...)
    for (i=0; i<path->n_points; ) {
       szT[0] = 0;
       switch (path->tags[i]) {
            case GF_PATH_CURVE_ON:
            case GF_PATH_CLOSE:       (...); i++;  break;
            case GF_PATH_CURVE_CONIC: (...); i+=2;  break;
            case GF_PATH_CURVE_CUBIC: (...); i+=3;  break;
    }   }    (...) }
\end{lstlisting}  \vspace{-5pt}
\hrule \vspace{-5pt}
\begin{lstlisting}[numbers=left,basicstyle=\small]
static int ftdi_elan_read_config(struct usb_ftdi *ftdi,  int config_offset,
                                 u8 width, u32 *data)
{(...)
  wait:
     if (ftdi->disconnected > 0) { return -ENODEV; }
     else {
        if (command_size < COMMAND_SIZE && respond_size < RESPOND_SIZE){
           ftdi_elan_kick_command_queue(ftdi);
           wait_for_completion(&respond->wait_completion);
           return result;
        } else {  msleep(100);  goto wait;  }
}    }
\end{lstlisting} \vspace{-5pt}
\hrule \vspace{-5pt}
\begin{lstlisting}[numbers=left,basicstyle=\small]
static int read_byte(struct file *file) {
   int ch = getc(file->file);
   if (ch >= 0 && ch <= 255){ ++(file->read_count); return ch;
   }  else if (errno == EINTR)  { /* Interrupted, try again */
       errno = 0;   return read_byte(file); 
   }  (...) }
\end{lstlisting} \vspace{-5pt}
\hrule \vspace{-5pt}
\caption{A \emph{loop} divergence in MP4Box (above); a \emph{goto} divergence in the FTDI network driver of the Linux kernel 5.19.1 (middle); a \emph{recursive} divergence in libpng (below).}
\vspace{-10pt}
\label{fig:examples}
\end{figure}
%\end{center}

\paragraph{Divergent loop} MP4Box is a large multimedia toolkit of almost a million lines of C code and 2900 stars on GitHub. MP4Box contains a number of of divergence issues, including one in the   \ \texttt{svg\_dump\_path} function as presented at the top of \cref{fig:examples}. 
In this function, a loop iterates over the points in an SVG path object. The loop contains a switch statement testing each path tag in the tag array for this point (line 6). 
As variable \texttt{i} is not incremented in the \texttt{for} loop and the \texttt{switch} statement does not include a default case,  the \texttt{for} loop divereges when there is a tag value other than GF\_PATH\_CURVE\_ON, GF\_PATH\_CLOSE, GF\_PATH\_CURVE\_CONIC or GF\_PATH\_CURVE\_CUBIC (the four cases). 
A possible fix is to add a default case to the \texttt{switch} statement such that all other tags return an error value (\eg \texttt{null}) to the upstream caller.

\paragraph{Divergent \texttt{goto}} The FTDI Linux kernel driver is a popular network driver which contains a number of non-terminating conditions due to handling \texttt{goto} statements incorrectly. One such issue is shown in the middle of \cref{fig:examples}, where a \texttt{wait} label on line 4 attempts to restart the logic of the \texttt{ftdi\_elan\_read\_config} function when the amount of the data read does not pass the \texttt{if} checks on line 7 (\eg when \texttt{RESPOND\_SIZE}=0). Subsequently, the \texttt{else} branch on line 11 is taken where the driver waits for 100ms and attempts to reread data. (This exact pattern is also used in other functions of the FTDI driver, \eg in the \ \texttt{read\_byteftdi\_elan\_read\_reg}, \texttt{ftdi\_elan\_read\_pcimem}, \texttt{ftdi\_elan\_write\_reg}, \texttt{ftdi\_elan\_write\_pcimem} and \texttt{ftdi\_elan\_write\_config} functions.)
%\azalea{It is not clear where the bug is here. Is it because they don't reread new data and keep checking the old stuff every time with the old data? Or is the assumption is that even if you read new data every time you could still fail the if case every time and diverge?}

%\begin{figure}[t]
%\hrule
%\small
%\begin{lstlisting}
%static int ftdi_elan_read_config(struct usb_ftdi *ftdi, 
%                                 int config_offset,
%                                 u8 width, u32 *data)
%{
%    (...)
%wait:
%     if (ftdi->disconnected > 0) {
%        return -ENODEV;
%     }  
%     else {
%        if (command_size < COMMAND_SIZE && respond_size < RESPOND_SIZE)
%        {
%           ftdi_elan_kick_command_queue(ftdi);
%           wait_for_completion(&respond->wait_completion);
%           return result;
%         } else {
%            msleep(100);
%            goto wait;
%         }
%      }
%}
%\end{lstlisting}
%\hrule
%\caption{A \emph{goto} divergence in the FTDI network driver of the Linux kernel 5.19.1.}
%\label{tab:ex2}
%\end{figure}

\paragraph{Divergent recursion} The \texttt{libpng} library is the official Portable Network Graphics (PNG) reference library, widely used as either a standalone library or often embedded in other project distributions.%, with over 1300 stars on GitHub. 
It contains non-termination conditions where the interruption of a system call within the \texttt{read\_byte} function leads it to call itself recursively and attempt to reread from the same file without any limit on the number of such attempts. This is shown at the bottom of \cref{fig:examples}, where the call to \texttt{getc} on line 2 attempts to get the next byte from the input file. However, when an EINTR error is detected (the \texttt{else if} branch line 4), \texttt{read\_byte} is called recursively (line 5) without any additional termination control. The recommended \emph{fuel} pattern would ensure that the number of attempts is finite (typically small) so that any abnormal interruption pattern is gracefully handled instead of trying to execute the same logic again without a forced exit condition.

%\begin{center}
%\begin{table}[]
%\begin{tabular}{|c|}
%\hline
%\small
%\begin{lstlisting}
%static int read_byte(struct file *file) 
%{
%   int ch = getc(file->file);
%   if (ch >= 0 && ch <= 255)
%   {
%      ++(file->read_count);
%      return ch;
%   }
%   else if (ch != EOF)
%   {
%      file->status_code |= INTERNAL_ERROR;
%      file->read_errno = ERANGE; /* out of range character */
%      /* This is very unexpected; an error message is always output: */
%      emit_error(file, UNEXPECTED_ERROR_CODE, "file read");
%   }
%  else if (errno == EINTR) /* Interrupted, try again */
%  {
%       errno = 0;
%       return read_byte(file);
%   }
%  (...)
%}
%\end{lstlisting} \\
%\hline
%\end{tabular}
%\caption{A \emph{recursive} divergence in libpng.}
%\label{tab:ex3}
%\end{table}
%\end{center}

\section{Design Of \pulseinf}
\label{sec:design}

We present the design of \pulseinf, the non-termination checker we have implemented over \pulse (which is built over the \infert framework). \pulseinf is the first tool implementing divergence detection following the under-approximate characterisation of \unter~\cite{UNTER}. 
%, guaranteeing \emph{no false positives}. 
\pulseinf can analyse very large bodies of code in a reasonable amount of time (see \cref{sec:evaluation}), leveraging the separation logic prover in the \pulse checker that underpins it. There are two main components of this prover: the first proves divergence in loops (and \texttt{goto} statements),  the second proves divergence in recursive functions.

\subsection{Loop (and \texttt{goto}) Divergence Detection}

We focus on our \pulseinf extensions over the \infert framework. The main idea is to detect loops where the accumulated path condition and the syntactic loop conditions repeat after executing the loop body a number of times (usually just once). Detecting this requires the following changes to \pulse.
%Our extensions are mostly centred around the four components below.
%: the abstract interpreter (\texttt{AbstractInterpreter.ml}), domain-specific additions (\texttt{PulseExecutionDomain.ml}) and the built-in solver (\texttt{PulseFormula.ml}).

%% \paragraph{Transfer functions} The original \pulse workflow uses a loop in which every instruction of the intermediate language, \emph{SIL} (Smallfoot~\cite{smallfoot} Intermediate Language), is symbolically executed on the abstract state based on the abstract semantics of the instruction (its \emph{transfer function}~\cite{absint}).
%% The \emph{path condition} is extended every time \pulse executes a conditional jump or enters a loop. 
%% %Adding new facts to the path condition triggers normalisation steps in the solver and conditions that are always true or false may get simplified away. 
%% We thus modify the solver to retain termination-sensitive constraints (see `the formula solver' below). % even when they can be simplified away.

\begin{figure}[t]
\footnotesize
\begin{lstlisting}
optim(int p) { int i = 0; while (i < 20) p++; }
non_optim(int i, int p) { while (i < 20) p++; }
loop_cond_nonterm(int y) { 
  int x = 0; while (y < 100) { if (y < 50) x++; else y++; } }
loop_pointer_nonterm(int *x, int y) { 
  int *z = x; if (x == &y) { while (y < 100) { y++; (*z)--; } } }
\end{lstlisting}  \vspace{-5pt}
\hrule \vspace{-5pt}
    \caption{Loop divergence examples detected by \pulseinf.}
    \vspace*{-15pt}
    \label{fig:loopexamples}
\end{figure}

\paragraph{The formula solver} \pulse uses its own custom-built SMT solver for reasoning about path conditions. This helps resolve SMT queries in a predictable amount of time. 
The internal representation of path conditions is always \emph{normalised} so as to 
\begin{enumerate*}[label=\alph*)]
	\item discover unsatisfiable paths as soon as possible; and 
	\item keep formulas compact. 
\end{enumerate*}

The existing solver cannot keep loop conditions as we found them given this normalisation, so we track these separately as a new addition. These \emph{termination conditions} are stored without being first normalised according to the current facts in the formula (which could, \eg discover that they are always true and discard them altogether). For example, the loop condition in function \texttt{optim} of \cref{fig:loopexamples} would be optimised away given that $x$ is initialized to $0$ and $0 < 20$ is always \texttt{true}, which would make the recurring state hidden due to the formula simplification. This issue would not appear in \texttt{non\_optim} given that \texttt{i} is not assigned a constant value within the function.
%% However, the solver was not adapted to detect divergence which requires it to keep all constraints influencing the path conditions of the program, even when these constraints can be \emph{optimised away} (\eg when they are always true). We refer to such optimised-away constraints as \emph{termination conditions} and have modified the solver to retain them as part of the state.

Other state values (variables other than those in path conditions and termination conditions) are discarded as they do not directly impact the control flow of the program under analysis. Retaining these dropped constraints would otherwise introduce false negatives, as irrelevant variable updates would prevent us from detecting the lasso. For example, updates to variable $p$ in \texttt{optim} and \texttt{non\_optim} would hide the lasso. Forgetting constraints is sound according to the proof system of \unter, which supports weakening post-conditions.

\paragraph{The abstract interpreter} \infert provides a generic abstract interpretation framework,  \inferai, that orchestrates its intra-procedural analysis. In particular, \inferai invokes the \emph{widening} operator of the abstract domain every time a loop header is reached. 
The \pulse widening is set up to converge after $k$ times (by picking whatever state we have have that step as the result of the widening), where $k$ is configurable and currently defaults to 3, making it an under-approximate finite loop unrolling.
We extend the widening operator to check for \emph{lassos}: a back-edge to a previously visited state with the same path condition (\ie accumulated list of conditionals) and termination conditions.

When such a lasso is found, we can report an \emph{infinite loop} error.
\pulseinf is able to detect divergence in examples \texttt{loop\_cond\_nonterm} and \texttt{loop\_pointer\_nonterm} in \cref{fig:loopexamples} with this mechanism. 
This logic is explained visually in more details in Appendix~\ref{sec:loopfig} (\cref{fig:design}).

We can also record such errors as part of the procedure's summary (in a new kind of pre/post tagged with \textsf{InfiniteProgram}) so that they are propagated to call sites, provided that the program path leading to the divergent loop is still feasible in the caller. In doing so, we can discover which infinite loops are \emph{reachable from an entry point}, which helps prioritise the list of issues. We used this technique to triage issues in the Linux kernel in \cref{sec:evaluation}.

%% \paragraph{The execution domain} We extend \pulse with a new type of state, \emph{InfiniteProgram}, to model states which have been determined as causing a divergence, and which are inserted in the state queue whenever a \emph{lasso} (a back-edge on a previously visited state) is detected. 
%% To do this, we identify the relevant part of the state as the location of the analysed loop, the current path condition, and the current termination conditions of the state. 

\subsection{Recursive Function Divergence Detection}

In addition to divergent loops, \pulseinf also flags divergent (mutually) recursive procedure calls. It does so by detecting when a function calls itself recursively (possibly via some other functions called) \emph{with the same state as its precondition}, in particular with the same values being passed as arguments to the call. This simple idea proved surprisingly effective in practice, especially on Hack code where complex call resolution algorithms in the runtime can catch developers off guard.  We now detail how this is implemented in \pulseinf.

\paragraph{Inter-procedural analysis scheduling in \infert} Let us start with how \infert performs \emph{compositional, inter-procedural analysis}. 
Each procedure in a given codebase is analysed independently (in isolation) by each ``checker'' in \infert (where \pulse and \pulseinf are such checkers). \inferai drives the intra-procedural analysis for each procedure and each checker. Checkers with inter-procedural capabilities compute a \emph{summary} for each procedure that is then stored in a database. The global analysis of the codebase is orchestrated as follows by a module called ``\ondemand'' inside \infert.

When the analysis of a procedure $f$ starts, we first add $f$ to a list of ``active'' procedures (those under analysis).  If, while analysing $f$, the current checker requests the summary of another procedure $g$ (typically in order to resolve a call to $g$ within $f$), then one of the following happens:
\begin{enumerate}
\item
  A summary for $g$ is found in the summary database and the analysis of $f$ can use it immediately.
\item
  No summary for $g$ is found.; subsequently we continue with computing a summary for $g$.
\begin{enumerate}
\item
  If we do not have the code for $g$ (\eg when $g$ is within a proprietary library), we return \texttt{UnknownProcedure} and let the checker apply its heuristic for unknown procedures.
\item
  If we do have the code for $g$, we check if it is \emph{active}. If so, we return \texttt{MutualRecursion} and let the checker apply its heuristic for recursive calls. This is new in \pulseinf.
\item
  Otherwise ($g$ is inactive and we have its code), we mark $g$
  active and begin analysing $g$. Once finished, we store $g$'s
  summary in the database and pass it back to the analysis
  of $f$.
\end{enumerate}
\end{enumerate}

\paragraph{Tracking recursive cycles in \pulseinf}
At step 2(b) above, \pulseinf records a special information in its abstract state that $g$ was a recursive call, together with the abstract values of the arguments to $g$ at the call site. 
That information is propagated in the summaries of callers of $f$, together with the sequence of such callers, and the abstract values eventually passed down to $g$ are updated each time to what they correspond to in each caller's summary (see the example below). If we reach $g$ again and apply a callee summary containing the recursive call to $g$ (which is bound to happen unless the call chain from $g$ to $f$ and back to $g$ is detected by \pulseinf to be infeasible), we can compare the sub-state rooted at the abstract values that will eventually be passed to $g$ by that call with the current precondition of $g$. If they are equal, we know we have found an abstract state $P$ (the precondition of $g$) such that $g$ is called in a context where $P \sep \textsf{true}$ holds, establishing a divergent behaviour in the abstract state from $P \!\sep\! \textsf{true}$ to $P \!\sep\! \textsf{true}$ by the \emph{framing principle} of separation logic.

Note that for the above scenario to happen at all, either we must have started analysing $g$ before it triggered the analysis of $f$, possibly via some other calls, or $g$ and $f$ are the same procedure.

%% One peculiarity of finding mutual recursion cycles via a compositional analysis is that the cycle is first discovered by the \textsf{Ondemand} module when going down the call graph, and then it is propagated back up by \pulseinf, from callee summary to caller summary, until the head of the cycle is reached.
%% \azalea{@Jules, why is this peculiar? :)}

\begin{figure}[t]
\footnotesize
\begin{lstlisting}
void trivial(int x) { trivial(x); }

void f(int *x, int *y) {
  int* z = (int*) malloc(sizeof(int)); if (z) { g(x, y); free(z); } }
void g(int* p, int* q) { if (*p > *q) { h(q, p); } }
void h(int* u, int* v) { f(v, u); }
\end{lstlisting}  \vspace{-5pt}
\hrule \vspace{-5pt}
    \caption{Two examples of divergent recursion flagged by \pulseinf.}
    \vspace*{-15pt}
    \label{fig:recexamples}
\end{figure}

\paragraph{Examples} Let us unroll this technique on the two strongly connected components of the call graph of \cref{fig:recexamples}. 
Let us first consider the \texttt{trivial} example: \pulseinf starts analysing \texttt{trivial(x)} by assigning a logical variable $x'$ to the original variable of $\texttt{x}$, yielding the precondition and current state $\texttt{x}=x'$. It then requests a summary for \texttt{trivial}. Since this is the currently active procedure, it gets back \texttt{MutualRecursion} back from \ondemand and records a recursive call to $\texttt{trivial}(x')$. This being a self-cycle, it is indeed trivial to check that $x' = x'$ and report an infinite recursion.

The cycle \texttt{f}, \texttt{g}, \texttt{h} in the second example is more interesting. 
Let us suppose that the analysis starts by analysing \texttt{f}. We get to the call to \texttt{g}; at this point the currently-inferred precondition is $\phi = \texttt{x}=x' \sep \texttt{y}=y'$ (where $\sep$ is the separating conjunction of separation logic, equivalent to $\wedge$ on pure predicates) and the current state is $\psi = \phi \sep  \texttt{z}=z' \sep z'\mapsto -$. 
The summary of \texttt{g} is as-of-yet missing, so \ondemand schedules its analysis, which, in turn and skipping slightly ahead, triggers the analysis of \texttt{h}. Requesting the summary for \texttt{f} finally yields a \texttt{MutualRecursion} reply from \ondemand since the active procedures are \texttt{f}, \texttt{g}, \texttt{h} at this point, which gets recorded in the summary of \texttt{h} as precondition $\texttt{u}=u' \wedge \texttt{v}=v'$ and $\textsf{RecursiveCall}(\texttt{f}, v', u')$.

The analysis of \texttt{g} resumes and eventually produces the precondition $\texttt{p}=p' \sep \texttt{q}=q' \sep p'\mapsto v_p \sep q'\mapsto v_q \sep v_p > v_q$ and  $\textsf{RecursiveCall}(\texttt{f}, p', q')$. 
(Note that here we omit the second disjunct where the comparison $v_p > v_q$ does not hold, and elide all post-conditions since they are irrelevant to our reasoning.)
This is because applying the summary for \texttt{h} yielded the substitution $u'\rightarrow q',\,v'\rightarrow p'$, which was applied to the recursive call predicate.

Finally, the analysis of \texttt{f} picks up the summary of \texttt{h} and applies its precondition successfully, yielding an updated precondition for \texttt{f}, namely $\phi' = \texttt{x}=x' \sep \texttt{y}=y' \sep x'\mapsto v_x \sep y'\mapsto v_y \sep v_x > v_y$, and a substitution containing $p'\rightarrow x',\,q'\rightarrow y'$. 
After applying this substitution, we obtain $\textsf{RecursiveCall}(\texttt{f}, x', y')$. 
The recursive call \texttt{f} is the same as the current procedure so a cycle has been found! 
\pulseinf now checks that the values $(x',y')$ passed to the recursive call are the same ones we started with, and that the sub-heaps rooted at those values are identical between the precondition and the current state, which is $\phi' \sep \texttt{z}=z' \sep z'\mapsto -$. The cycle is thus divergent, leading either to an infinite recursion, a stack overflow, or undefined behaviour.

\paragraph{Non-determinism of results}
The astute reader may wonder whether the order in which procedures are analysed can influence the results. In general, the order does matter and can cause false negatives (\ie missing divergence bugs, in keeping with the spirit of under-approximation which prioritises no false positives at the expense of false negatives). The technique being largely incomplete, this is just another source of incompleteness (\ie under-approximation). 
\infert will always analyse mutually-recursive cycles in the same order so the analysis still produces deterministic results.

%% \begin{figure}[t]
%%     \centering
%% \includegraphics[width=\textwidth]{pulse-infiniterec.png}
%% 	\vspace{-20pt}
%%     \caption{Design of the infinite recursion checker.}
%%     \label{fig:design}
%% \end{figure}

\section{Evaluation}
\label{sec:evaluation}

We report on a fully reproducible evaluation of our divergence verifier \pulseinf. We have identified new vulnerabilities in each of the categories of divergence previously presented (\texttt{goto} loops, infinite loops and infinite recursions). We report on the size and runtime associated with each analysed component in \cref{tab:eval1}. Our reproducible experiments on open source projects were performed on a single Dell PowerEdge R7515 server equipped with an AMD EPYC 7543P processor of 32 cores and 2 GHz clock.

\begin{table}[t!]
    \centering
    \begin{tabular}{|c|c|c|c|}
\hline
\textbf{Analysed Project} &	\textbf{Prog Lang} & \textbf{Analysed LOC\#} & \textbf{Analysis Time} \\
\hline
OpenSSL	        & C   &	804K &	1m26  \\
libpng	        & C   &	96K &	6s \\
zlib	        & C   &	41K &	7s \\
libpcre2	    & C   &	133K &	18s \\
libxml2	        & C   &	300K &	57s \\
mbedTLS	        & C   &	554K &	25s \\
CryptoPP        & \C++ &	51K &	2m25 \\
libxpm	        & C	  & 11K &	2s \\
Lua	            & C	  & 30K &	22s \\
LibGit2	        & C	  & 374K &	29s \\
Open5GS	        & C	  & 1.5M &	1m4 \\
FreeImage	    & \C++ & 461K &	12s \\
Bitcoin Core	& \C++ & 250K &	4m54 \\
Comdb2	        & C   & 856K &	1m49 \\
BlazingMQ	    & \C++ & 6.5M &	7m24 \\
BDE	            & \C++ & 4.03M &	1m44 \\
MP4Box (gpac)	& C	  & 911K &	45m \\
Linux kernel (5.19.1) &	C &	26M &	10m35 \\
SQLite	        & C   &	446K &	3m41 \\
Wireshark	    & C   &	5.6M &	6m25 \\
bind	        & C   & 463K &	20s \\
ProFTPD	        & C   &	356K &	1m12 \\
Exim	        & C   &	335K &	16s \\
\hline 
\textbf{TOTAL}	 & &	50.1M & \\	
\hline
    \end{tabular}\vspace{5pt}
    \caption{Performance of \pulseinf on 50 million lines of reviewed projects.}
    \vspace*{-20pt}
    \label{tab:eval1}
\end{table}

\paragraph{Scalability} \pulseinf boasts impressive scalability, analysing million-line projects in just over one minute. \pulseinf was able to analyse the entirety of the Linux kernel (26M LoC) in just over 10 minutes. To our knowledge, this is the first ever divergence checker capable of scaling to such extent. Additionally, our techniques have been run on over 100 millions of proprietary industrial code, generating hundreds of alerts. This illustrates the extreme scalability of our method. A small sampling of the alerts has revealed numerous true positives, some of which have already been fixed. However, we only report triage results for the 50M LoC open source code in this paper, which can be independently assessed through the artifact associated with the paper.

\paragraph{Bugs found} \pulseinf has identified more than 30 unintended divergences in open source programs as listed in \cref{tab:evalbugs}. Our findings have been shared with developers and several issues have already been fixed. Identified issues include many cases where the program calls into a function which is allowed to fail and later restarted. This is the case for calls to \texttt{malloc} (which returns \texttt{null} in case of failure) and I/O syscalls such as read and write which can be interrupted and return EINTR to signal that it should be called again. Other cases include network endpoints where a minimum amount of data is expected and the program will wait until it has read all the data needed without guaranteeing termination.

\paragraph{False positives} The theory of \unter~\cite{UNTER} guarantees \emph{no false positives}, but only relative to environmental assumptions such as concerning library code:  Like other symbolic execution tools, \pulseinf can have false positives for these reasons. For example, a divergence may depend on the result of a library function we do not analyse, or may depend on some initial conditions that are not captured by the inter-procedural analyser. In our experience many of these false positives can be eliminated by manually adding missing API models, a feature already well-supported by \pulse. In practice, the number of false positives is very small given the scale of our experiments, and reviewing them manually was practical.

\begin{table}[t]
    \centering
    \begin{tabular}{|c|c|c|c|c|c|}
\hline
\textbf{Analysed project} &	\textbf{Recursive alerts} & \textbf{Recursive bugs} & \textbf{Loop alerts}   & \textbf{Loop bugs} & \textbf{Intended infinite} \\
%\textbf{Project} &	\textbf{Alerts}    & \textbf{Bugs}      & \textbf{Alerts} & \textbf{Bugs} & \textbf{Infinite} \\
\hline
OpenSSL	        &  0 & 0 &  12 & 4 & 3	\\
libpng	                  &  1 & 1 &   2  & 0   & 0 	\\
zlib	                  &  0 & 0 &    2  & 1   & 0  \\
libpcre2	       &  0 & 0 &   0  & 0   & 0  \\
libxml2	        &  1 & 0 &    9 & 5   & 0  \\
mbedTLS	        &  0 & 0 &     2 & 0   & 0 	 \\
CryptoPP              &  0 & 0 &     6 & 2  & 1  	\\
libxpm	                  &  0 & 0 &     0 & 0  & 0  \\
Lua	                  &  0 & 0 &     2 & 0  & 0 	 \\
LibGit2	        &  0 & 0 &     11 & 4  & 0  \\
Open5GS	        &  0 & 0 &     5 & 0  & 0 	 \\
FreeImage	       &  1 & 0 &     22 & 3  & 0  	 \\
Bitcoin Core	       &  0 & 0 &     2 & 0   & 0  	 \\
Comdb2	       &  0 & 0 &    48 & 3  & 7  \\
BlazingMQ	       &  0 & 0 &     7 & 0  & 4 	 \\
BDE	                  &  0 & 0 &     17 & 0  & 12  	 \\
MP4Box (gpac)    & 0 & 0 &     18 & 1   & 3  	 \\
Linux kernel         & 3 & 0 &     19 & 7  & 2  	 \\
SQLite	                 & 0 & 0 &     6 & 0   & 0   \\
Wireshark	      & 0 & 0  &     63 & 3  & 0  \\
bind	                 & 1 & 0 &     4 & 0   & 1  \\
ProFTPD	      & 0 & 0	&     13 & 2   & 3  \\
Exim	                 & 0 & 0	&   10  &  3   &  0  \\
\hline 
\textbf{TOTAL}	 & 7 &	1 &  280 & 38  & 36 \\	
\hline
    \end{tabular}
    \vspace{5pt}
    \caption{Findings for goto, loop, and recursive non-terminations (excluding test code).}
    \vspace*{-20pt}
    \label{tab:evalbugs}
\end{table}

\paragraph{Intended divergence} Several programs such as OpenSSL, ProFTPD or the Linux kernel, include a number of intended divergence code patterns, which are correctly detected by our tool and classified as such. Instances of intended divergence include cases where a thread is created whose role is to pump events from a queue and wait until the next event pops up if the queue is empty. A similar pattern arises when calling the \texttt{scanf} or \texttt{readline} functions that wait on user-controlled events to unblock, which we consider intended divergence. Another common uncovered pattern happens when a program attempts to lock a mutex and will wait until the mutex is released by another thread. In absence of support for concurrent non-termination proving in \unter and thus \pulseinf, we classify such patterns as intended divergence.

\paragraph{In-situ evaluation: infinite recursion} The \pulseinf checker for infinite recursion has been analysing Meta developers' code changes at code-review time as part of a continuous integration (CI) system for several weeks, in one of Meta's largest codebases: its Hack codebase. The Hack codebase at Meta has hundreds of millions of lines of code. In just a few weeks of being deployed, \pulseinf has prevented tens of divergent recursive functions to be committed to the codebase and potentially run in production, with virtually no false positives. The issues are not limited to recursive methods, which are the most common but so far represent less than half of all issues; mutually-recursive methods are the majority and typically involve two or three methods per cycle.

\section{Related Work}
\label{sec:related}

There is a large body of work on tools for proving program termination for industrial programs. Terminator~\cite{cook2006terminator} and its successor T2~\cite{brockschmidt2016t2} are some of the earliest termination provers capable of handling pointers, nested loops and non-determinism. Terminator was deployed on Windows kernel drivers whose implementation is not publicly available. Terminator and T2 also only support proving termination rather than non-termination. These tools use SMT solving to implement an over-approximate termination prover, leading to a number of false positives. Cooperating-T2 is an improved variant of the Terminator family, whose performance is improved but still suffers from the same caveats as other versions. HIP-TNT~\cite{le2014resource} and HIP-TNT+~\cite{le2017hiptnt+} further improved this approach by adding support for separation logic to Terminator; however they do not support under-approximation or basic data structures such as arrays, limiting their applicability to small programs.

Mutant~\cite{berdine2006automatic} is the first termination prover leveraging separation logic, however it was only applied to simple programs in a toy language. The first non-termination prover capable of analysing programs is the work of Gupta \etal~\cite{gupta2008proving} which defines the concept of a recurrent set which can be used to prove non termination of a buggy binary tree sorting implementation. They did not analyse any large programs or use separation logic for compositional analysis. But, their concept of recurrent set is fundamental and is a close relative of the idea of a repeating state in an under-approximate abstract semantics. 

Key~\cite{velroyen2008non} is another tool capable of proving non-termination, however it only analyses \emph{integer C programs} (with only integer datatypes, without pointers and without function calls), and was only evaluated on 55 test programs. 
CPROVER~\cite{kroening2010termination} is a model checker by Kroening \etal capable of proving termination for Windows drivers using k-bounded analysis; however, it does not include support for separation logic or heap programs. 
\textsc{Caber}~\cite{caber} is a tool for proving termination (not divergence), and while it supports heaps, it has only been applied to a handful of small programs,  and not large codebases or libraries. 
Many existing tools \cite{brockschmidt2013better,larraz2014proving,chatterjee2021proving} can also \emph{only} handle integer C programs, and thus, unlike \pulseinf, they \emph{cannot} run on existing C codebases or libraries such as \openssl, unless they are first pre-processed into integer C programs.

DynamiTe~\cite{le2020dynamite} and AProVE~\cite{hensel2022aprove} are some of the latest tools capable of proving termination as well as non-termination using SMT and recurring set analysis. Our tests show that our results are comparable to DynamiTe on SV-COMP non-linear arithmetic benchmarks~\cite{UNTER}. While providing good results on benchmarks, our discussion with the authors of these tools confirmed that they were never deployed on any real world program. These tools also required the entire target program with a main function and were unable to analyse libraries or incomplete programs.

\section{Artifact}
\label{sec:artifact}
Source code and documentation for \pulseinf are available at \url{https://github.com/jvanegue/infer/}.

\bibliographystyle{splncs04}
\bibliography{biblio}

\paragraph{Acknowledgements}
%We would like to thank the CAV reviewers for their insightful and constructive feedback. 
Raad is supported by the UKRI Future Leaders Fellowship MR/V024299/1, by the EPSRC grant EP/X037029/1, and by VeTSS.
\clearpage
\appendix
\section{Loop Divergence Detection Logic}
\label{sec:loopfig}

\begin{figure}[h]
    \centering
    \includegraphics[width=\textwidth]{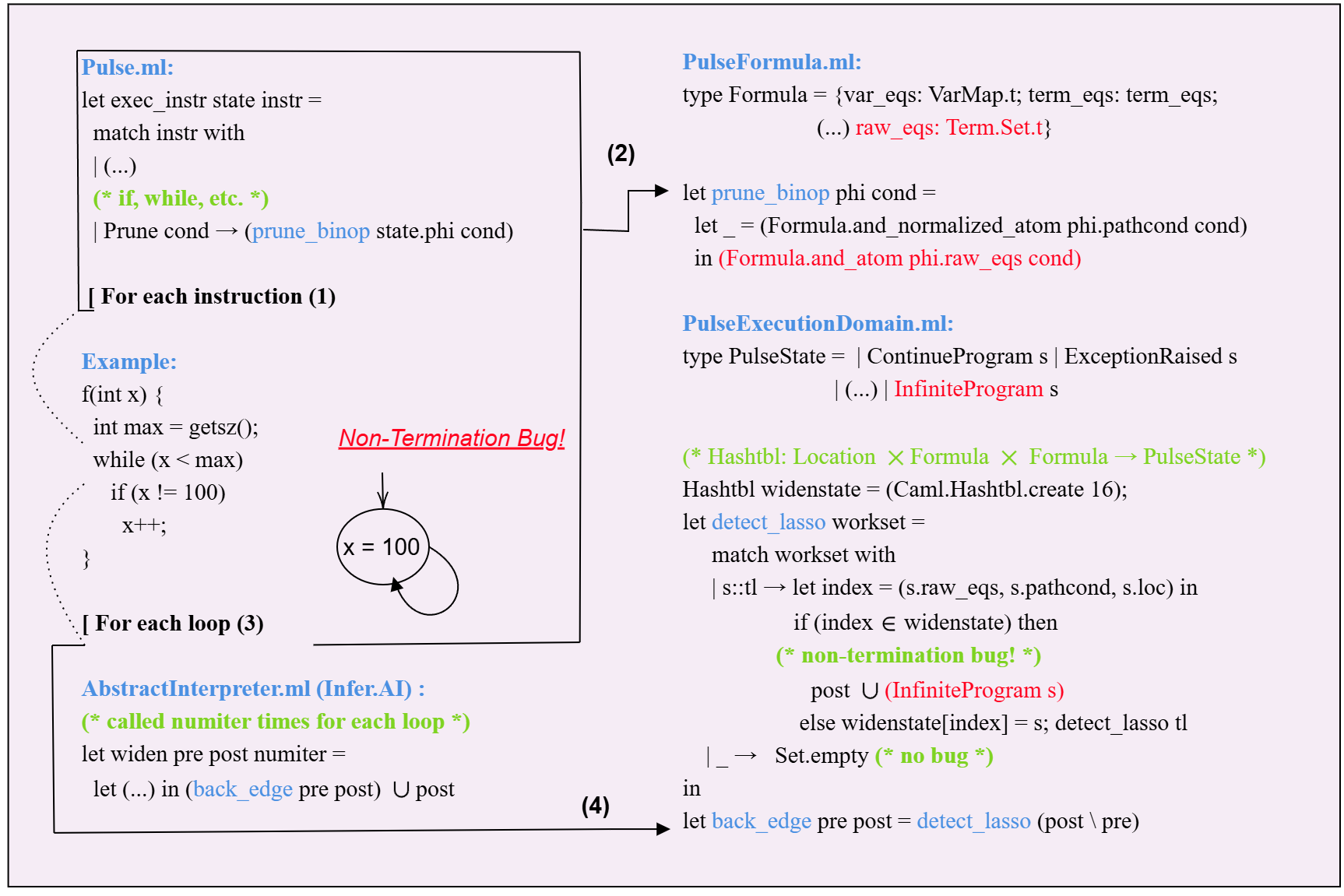}
    \vspace{-20pt}
    \caption{Design of the infinite loop checker.}
    \label{fig:design}
    \vspace{-10pt}
\end{figure}

\section{Divergence Vulnerabilities}
\label{sec:cve}

%
%\begin{wrapfigure}{r}{0.45\textwidth}
%\vspace{-13pt}
%\centering
%\begin{tikzpicture}[scale=0.6]
%\label{cvetrend}
%\begin{axis}[
%%    title={Vulnerability trend for divergent behavior bugs},
%    xlabel={Year},
%    ylabel={Number of published bugs},
%    xmin=1999, xmax=2022,
%    ymin=0, ymax=120,
%    xtick={1999,2002,2005,2008,2011,2014,2017,2020,2023},
%    ytick={0,20,40,60,80,100,120},
%    legend pos=north west,
%    ymajorgrids=true,
%    grid style=dashed,
%]
%\addplot[color=blue, mark=square]
%coordinates {
%(1999,1)(2000,4)(2001,0)(2002,9)(2003,10)(2004,30)(2005,47)(2006,34)(2007,60)(2008,26)(2009,31)(2010,35)(2011,26)(2012,39)(2013,58)(2014,38)(2015,46)(2016,57)(2017,103)(2018,92)(2019,43)(2020,56)(2021,43)(2022,28)
% };
%\legend{Number of Divergence CVEs per year}    
%\end{axis}
%\end{tikzpicture}\vspace{-10pt}
%%\end{center}
%\caption{Vulnerability trend for divergence bugs.}
%\vspace{-20pt}
%\label{fig:cve_trend}
%\end{wrapfigure}

Divergence bugs are widespread across a number of programming languages. We present several examples taken from the \href{https://www.cve.org/}{Common Vulnerabilities and Exposures} (CVE) database and categorise them along common cases of vulnerabilities. % -- see \cref{fig:cve_trend} for the prevalence of divergence bugs. 
%We first take a quantitative approach by looking at the bug trend over more than two decades of security research, as shown in \cref{fig:cve_trend}. 
%\Cref{fig:cve_trend} depicts the prevalence of vulnerabilities for divergence bugs.
We focus on control-flow-related divergent behaviours brought about on certain inputs. 
%control control flow program behavior leading to the program not terminating on some input. 
%Our under-approximate approach guides us towards incorrectness specifications that are guaranteed to yield no false positive (NFP). Capturing NFP heuristics for control-flow related bugs is a practical approach to tackle divergence detection, unlike unrestricted heuristics which can guarantee the absence of neither false positives nor false negatives.

In particular, we focus on capturing behaviours where non-termination is not intended (unlike interactive programs whose non-termination is expected and induced from an infinite message loop treating streams of incoming input requests), and guarantee that our approach focuses on detecting the most widespread vulnerability classes in publicly available code. We have selected a number of bugs that show a wide cross-section of programming languages and control flow conditions.

%\vspace*{-1pt}
\paragraph{Infinite Loops}
Recursive implementations are common in parsers. 
In some cases, the loop condition is driven by the value of an integer variable (\eg remaining stream bytes to be read),
%, size of next data unit to read on stream) 
which can be dynamically set within the parsing loop as the parser reads the input. 
If the decrement value of such variable in an iteration is set to 0, the loop makes no more progress leading to an unintended divergence. 
Specifically, when a parsing sub-function $f$ is called to treat a sub-case of input data type, if $f$ returns 0, then the loop makes no progress reading input. Such an example was found in the popular Wireshark network analyser, leading to \href{https://nvd.nist.gov/vuln/detail/CVE-2022-3190}{CVE-2022-3190}. 
% (see \cref{subapp:wireshark}). 

%\vspace*{-1pt}
\paragraph{Infinite Recursion}
Infinite recursion bugs are one of the main sources of divergence. 
%It is commonly known that every iterative loop can be encoded as a recursive function, which is a common implementation choice. 
Infinite recursion bugs are well-known to parser developers when the recursive parsing function allows input variable expansion or other generative capability, such that when the newly generated input after expanding variables is parsed through a recursive call, the number of subsequently needed recursive calls remains non-null. Such a case was seen in the widely used Log4j logging library for Java programs, leading to \href{https://nvd.nist.gov/vuln/detail/cve-2021-45105}{CVE 2021-45105}. 
% (see \cref{subapp:log4j}).

%\vspace*{-1pt}
\paragraph{Out-of-Order Transition Divergence}
Unintended divergence can result from a loop or recursive call to a parsing function where certain input values or record data types are expected to be treated in a certain order, and an out-of-order encoding results in an infinite cycle. In certain cases, special input tag types are intended to be found at certain parsing stages as to disallow spurious transitions.
% to other undefined states (including the current state itself). 
Such a vulnerability was discovered in the GraphQL language interpreter, where the \emph{string} type name can be encoded in the input such that the parsing handler calls itself repeatedly.
%(see \cref{subapp:golang} for an example vulnerability affecting Go programs).

%\vspace*{-1pt}
\paragraph{Zero-Sized Input Divergence}
Container data structures (\eg lists or vectors) are typically implemented with access primitives where adding or removing elements can be achieved independently of the current number of elements in the container. This is done by maintaining a meta-data size field. 
When such data structures are implemented with linear memory access in mind, an additional size field is needed to ascertain the size of an element in the data structure. 
Whether such element is of a fixed or variable size, an element with zero size can lead to a container iterator that diverges when traversing the structure without making progress. Such a problem was identified in the Linux kernel, leading to \href{CVE-2020-25641}{CVE-2020-25641} and was fixed in Linux kernel version 3.13. 
% (see \cref{subapp:kernel}).

%\vspace*{-1pt}
\paragraph{Offset-Encoded Divergence}
%A slice of parser programs is such that 
In parser programs it is sometimes possible for the input to describe the actual input offset at which the data object is found. 
When such input offset indirection occurs, a parsing loop or recursive function can diverge by returning to previously parsed input in a way that will redo previously completed work and diverge. An example of this bug can be found in the popular graphic software Blender, written in C. Additional state would be required to ensure that the current input offset is restored after such out-of-bounds element is read.
% (see \cref{subapp:blender}). 

%\vspace*{-1pt}
\paragraph{Exception-Induced Divergence}
Some parser implementations use exceptions to treat special or error cases where a recovery logic must be encoded in a catch or except block. 
%so that parsing can continue and ignore local errors. 
%, in a way that such error is recoverable and parsing can continue on the rest of the input. 
Exception-induced spurious transitions can then be encoded such that the induction variable is never incremented/decremented, leading to divergence. A particular example of such vulnerability can be found in the \emph{Sklearn} industry-standard library for machine learning and data analysis in Python, where a convergence-based discretisation algorithm can be made to never terminate if the exceptional execution path fails to break from the appropriate number of loop nesting levels.
% (see \cref{subapp:sklearn}).

%\vspace*{-1pt}
\paragraph{Algebraic Divergence}
Divergence bugs can be found in mathematical software, where specific algebraic conditions are expected on the input to reach a fixpoint in an iterative or recursive function. 
%, where an algebraic operation is performed on the input at each iteration, and the associated loop termination condition is defined as a numerical constraint over the input, or an input-dependent value. 
The \openssl cryptographic library contains such code, where a modular square root implementation for an elliptic curve group expects the residue of the recursive operation to reach value 1 eventually, but invalid input 
%curve 
parameters fail to meet this condition, leading to \href{https://nvd.nist.gov/vuln/detail/CVE-2022-0778}{CVE-2022-0778}. 
This vulnerability allowed remote SSL/TLS connections to get stuck in an infinite loop.
% (see \cref{subapp:openssl}). 
This example illustrates that even security code can be vulnerable to divergence bugs!

%% HERE WE CAN INCLUDE MORE DETAILS AND THE FIGURES THAT JULIEN CREATED TO EXPLAIN THE IMPLEMENTATION

%% \begin{figure}[t]
%%     \centering
%% \includegraphics[width=\textwidth]{pulse-infiniteloop.png}
%% 	\vspace*{-5pt}
%%     \caption{Design of the infinite loop checker. \jules{We should narrate the flow that is depicted in the figure here. Change \texttt{raw\_eqs} to \texttt{termination\_conditions} to match the text?}}
%%     \label{fig:design}
%% \end{figure}

%% \begin{figure}[t]
%%     \centering
%% \includegraphics[width=\textwidth]{pulse-infiniterec.png}
%% 	\vspace{-20pt}
%%     \caption{Design of the infinite recursion checker.}
%%     \label{fig:design}
%% \end{figure}

\end{document}